# Feasibility, Architecture and Cost Considerations of Using TVWS for Rural Internet Access in 5G

Mohsin Khalil[#], Junaid Qadir[*], Oluwakayode Onireti[$], Muhammad Ali Imran[$], Shahzad Younis[#]

National University of Sciences and Technology Pakistan[#], Information Technology University Lahore Pakistan[*], University of Glasgow UK[$]

mkhalil.msee15seecs@seecs.edu.pk, junaid.qadir@itu.edu.pk, oluwakayode.onireti@glasgow.ac.uk,
muhammad.imran@glasgow.ac.uk, muhammad.shahzad@seecs.edu.pk

*Abstract*—The cellular technology is mostly an urban technology that has been unable to serve rural areas well. This is because the traditional cellular models are not economical for areas with low user density and lesser revenues. In 5G cellular networks, the coverage dilemma is likely to remain the same, thus widening the rural-urban digital divide further. It is about time to identify the root cause that has hindered the rural technology growth and analyse the possible options in 5G architecture to address this issue. We advocate that it can only be accomplished in two phases by sequentially addressing economic viability followed by performance progression. We deliberate how various works in literature focus on the later stage of this 'two-phase' problem and are not feasible to implement in the first place. We propose the concept of TV band white space (TVWS) dovetailed with 5G infrastructure for rural coverage and show that it can yield cost-effectiveness from a service provider's perspective.

## I. INTRODUCTION

With the evolution of time, technological advancements have emphasized the need for new trends to cope up with the emerging requirements in every field. In case of wireless telecommunication, the prospect of meeting the ever-increasing requirements due to envisaged saturation of existing cellular capabilities calls for new directions for evolution. Over the years, the universal coverage aspect of cellular communication has remained unresolved akin to an insurmountable peak for four generations. This is because it is not economical for service providers to operate in the low Average Revenue per User (ARPU) regions, which has led to the formation of large coverage holes especially in the rural vicinity. The 5G networks are expected to be operational in near future and do not present promising solution for bridging the digital gap between rural and urban areas [1] – [3].

According to the 2015 statistics, it has been revealed that about 56% of the world population does not have access to the Internet. Hence in September 2015, United Nations marked the universal and affordable Internet provisioning to everyone everywhere by 2020 as a sustainable development goal [4]. In this context, the call for utilizing 5G for Global Access to Internet for All (GAIA) is need of the hour, since the current technological leads in this direction are not encouraging as of now. It is despite the fact that the areas with access to Internet result an increase in GDP growth as compared to those without Internet access [5]. In Singapore's case, Information and Communication Technology (ICT) contributed 1% GDP increase in its economic growth [6].

Since all the debate on technology provisioning is profit-driven, so it is feared that 5G has little to offer for the rural case. The mobile operators target residential patterns for connectivity, so the concept of 5G is assumed to be urban in nature. That is why, the vendors have translated the mantra of 'Coverage Everywhere' to 'Service Areas' (calculated via residential patterns) since they do not find any incentives due to low user density and lack of communication infrastructure in rural areas [3]. Moreover, the cost of extending Internet services with existing proposals in these areas is estimated to be higher than the expected revenue generation, therefore the rural population is deprived of Internet service due to lack of economic viability.

The research fraternity have come up with various solutions to address the aspect of universal coverage. The use of satellites and aerial platforms has been suggested to address this long-standing problem [7]. With the help of community networking, GSM white spaces can also be exploited for subject purpose [8]. Google has introduced the concept of using balloons for provisioning of Internet, in which about 300 balloons can cover up the earth's inhabited regions [9]. The use of white spaces in TV band of the spectrum has also been suggested to alleviate the issue in conventional cellular networks [10], but their use-case specific to 5G networks has not been advocated yet. Moreover, it is still unclear how these proposals might fit into the business model for rural coverage.

The main contribution of this paper is how low-cost rural Internet access can be accomplished in a 5G environment. We have explored root causes for this digital divide and have suggested to address this problem phase-wise by prioritizing availability over performance, since service availability is a more pressing concern than high performance (which will only be relevant when availability has been ensured). Furthermore, we have presented a network infrastructure model with lesser costs and simple architecture to make it feasible from a service provider's viewpoint. The deployment scenario of this model has also been formulated, where it is further transformed into an optimization problem for cost minimization to ascertain its practical viability. The comparison of this approach with conventional solutions presents encouraging returns in terms of cost savings.

This paper is organized into six sections. In order to make this paper self-contained, we succinctly describe a formative architecture of 5G network in Section II, based on which the case of rural Internet coverage can be exploited. In Section III, we explore the feasibility of rural coverage in 5G networks and show how various works proposed in existing literature can be found wanting in achieving this goal since they target availability and performance in a single phase. In Section IV, we present our vision to address the rural connectivity by utilizing the proposed concept of TVWS and how it may be slotted within existing 5G infrastructure. Our analysis for cost minimization in terms of Capital Expenditures (CAPEX) and Operational Expenditures (OPEX) is appended in Section V. The conclusion is annotated at the end in Section VI.

## II. FORMATIVE ARCHITECTURE OF 5G

The statistics on wireless usage indicate that on average, more than 70% of the data traffic is generated indoors [11]. The existing cellular designs make use of a single outdoor Base Station (BS) in a macrocell (located at the center) irrespective of indoor or outdoor connectivity. However, this arrangement might falter in future especially for indoor users due to ever-increasing throughput demand. It is because the penetration loss due to buildings/walls becomes significantly high in indoor environment which is detrimental to data rate and energy efficiency. For this purpose, the architecture of 5G is based on the revamping of existing cellular infrastructure. It proposes the separation of scenarios for outdoor and indoor connectivity in order to mitigate this penetration loss and to ensure enhanced network performance to indoor users.

The proposed 5G architecture is heterogeneous and can comprise of microcells, macrocells, relays and mobile femtocells in a Cloud Radio Access Network (C-RAN) in order to support promising wireless technologies such as massive Multi-Input Multi-Output (massive MIMO), Device-to-Device (D2D) communication, spatial modulation and millimeter wave (mmWave) communication [12], [13]. The 5G cellular network is proposed to be ultra-dense by employing these technologies. The mmWave Base Stations (MBS) are required to be deployed with greater density than macrocell base stations so that good coverage can be achieved. By moving towards mmWave spectrum, we can leverage more throughput, therefore ultra-densification motivates the positioning of multiple small cells in forthcoming cellular networks [14]. The option of transmitting the backhaul traffic of every MBS by Internet/fiber does not constitute an economic model. Additionally, since the small cell BSs employ mmWave communication, their transmission distance is significantly reduced due to greater path loss at higher frequencies as compared to microwave case. Therefore, the MBS cannot forward the backhaul traffic directly to the macrocell gateway and a distributed architecture is required for these ultra-dense topologies for relaying the traffic through multi-hop links.

In case of MBS, the inter-site distance similar to that of a microcell or picocell deployment can be utilized, whereas a macrocell can comprise multiple MBS as gateways based on the network topology. In order to take care of user mobility, the macrocell BS constitutes the Control Plane and MBS formulates the Data Plane [15]. The logical architecture of 5G is depicted in Figure 1 which shows multiple small cells within a macrocell. The base stations in the small cells (MBS) are connected through mmWave links. MBS 'B' and 'C' are acting as gateway due to their connectivity via Fiber-to-the-core (FTTC) link.

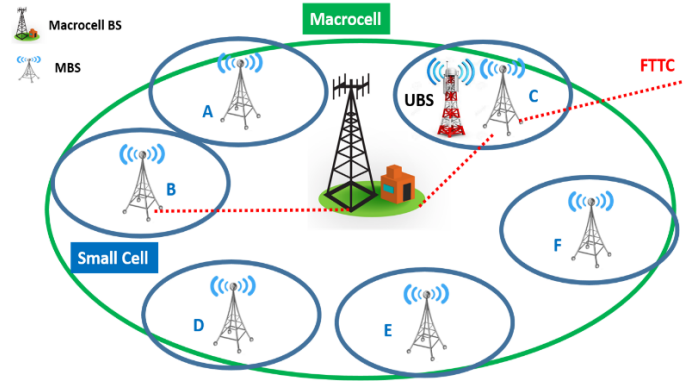

Fig. 1: Minimal 5G Architecture with two gateway MBS

## III. FEASIBILITY & CHALLENGES FOR RURAL COVERAGE

Although many other technologies such as beamforming and massive MIMO are also considered an integral part of envisaged 5G architecture, delving in those details is not required at this stage since we are focusing on the feasibility of rural Internet access. Since the majority of rural population would generally comprise of low-income customers, therefore provisioning of Internet in these areas with state of the art QoS guarantees would be an overambitious approach. The core problem is that the rural areas are deprived of Internet, so the foremost issue is that of availability instead of performance. Hence, this problem has to be addressed phase-wise. Phase-I embodies economic viability and Phase-II represents performance progression.

In the *first phase*, our aim is to ensure the availability of Internet in a cost-effective manner. This is only possible if the proposed model is able to attract the attention of service provider in the first place. By focusing on cost effectiveness, parameters such as throughput and latency can be compromised because these are managed according to the user requirements and are not deemed critical for a new user located at a remote/sparsely populated area. In addition, the infrastructure cost has also to be practicable because the vendor would never risk a huge sum for a pilot project. Since the profit-loss breakeven point would occur at lower revenues, therefore more rural population would be encouraged to reap the benefits of low-cost Internet. Once the rural access to Internet gains momentum and results an increase in number of users, the core problem would be addressed and would subsequently be wiped off the 'investment blacklist' from a service provider's perspective. In the *second phase*, the increased rural demand would motivate the service providers in facilitating the users with better connectivity and improved performance that would invariably lead to technological advancement in the rural areas and bridge this technology gap.

Various works in literature have suggested architectures for coverage of rural and low-income areas, but these are only feasible in Phase-II of the core problem. The proposal forwarded in [2] can only take off provided it offers money-making incentives for the vendor; hence the Phase-I of the problem remains unresolved. The use of satellites leverages ubiquity in the area under consideration, however its connectivity requires costly user equipment which renders it unsuitable for rural case. The idea of using GSM white spaces has also been floated in [8] but it would require the rural community to establish their own community networks. Terragraph project by Facebook [16] makes use of a multi-node 60 GHz wireless system for providing high speed Internet and helps in achieving street-level coverage of Gigabit Wi-Fi, but it is focused only towards dense urban areas. Moreover in terms of spectral efficiency, Facebook have launched ARIES project which employs 96 transmitting antennas in an array [16]. However, due to huge infrastructure costs, this approach cannot be considered as a solution for rural coverage. The use of Unmanned Aerial Vehicles (UAVs) and drones have surfaced as an aerial option for Internet connectivity, but this solution is only viable for metropolis and costs associated with their deployment in a village/hamlet are too high to afford [7]. The use of TV band white spaces has been tested for rural Internet connectivity by researchers in [17] with relatively lower costs, however the provisioning of optical fiber in villages for backhauling traffic might be a tough ask for thorough rural coverage.

## IV. USE OF TVWS WITH 5G ARCHITECTURE

Based on rural coverage challenges, it is safe to articulate that the rural Internet solution has to be centered on (i) scalable topology, (ii) minimum infrastructure deployment, and (iii) use of unlicensed band. In this context, the use of TV band white space presents itself as a promising technology since it fulfills all these requirements. Over the years, the researchers have carried out miscellaneous experiments to utilize unused TV UHF band for various communication scenarios. The use of TVWS as a networking solution has been materialized in a variety of cases with different preferences, a few of which are listed in Table I. It is pertinent to highlight that to the best of our knowledge, TVWS implementation with infrastructure cost-effectiveness as its core concern has not been proposed till date. Moreover for cellular networks, the most recent case of TVWS utilization has been demonstrated for Long Term Evolution (LTE) system [10]. In this case, an LTE base station makes use of TVWS and results have shown that downlink speeds up to 45 Mbps can be achieved successfully, albeit with specialized user equipment. The researchers have also advocated TVWS for rural connectivity by making use of optical fiber in selective villages to backhaul the aggregated data to a centralized point [17], however, this arrangement might not be an optimistic investment for a service provider on a large scale. In this context, we propose that the vista of using TVWS for rural Internet access can be utilized with the wireless backhaul of 5G architecture in a cost-effective manner. Since the 5G network makes use of C-RAN architecture due to resource pooling and centralization benefits, hence the Baseband Unit (BBU) of various BSs would be located at a centralized place. Therefore, the MBS cell sites would only house the Remote Radio Head (RRH) [18].

TABLE I
PRIMARY TARGETS IN VARIOUS USE-CASES OF TVWS

| Primary Focus | Concern-specific TVWS Use-case |
|---|---|
| Operational Feasibility | Test-bed for Internet connectivity [19] |
| Indoor Environment | Alternative for Wi-Fi in university campus [20] |
| Coverage Enhancement | TDMA Mesh network for rural Internet [21] |
| Spectrum Reuse | Hybrid system with smart utility networks [22] |
| Cloud Services | A localized hybrid TVWS Wi-Fi network [23] |
| Throughput | Operation and analysis of LTE System in TV band [10] |
| Infrastructure Cost | Not available |

It is pertinent to mention that a macrocell in a 5G environment can comprise of multiple MBS as gateways based on the network topology. The UHF Base Station (UBS) can be deployed at a site collocated with MBS. In addition to spectrum sensing, the UBS would also be connected to an online Spectrum Database Manager in order to quantize the availability of white spaces in the TV band for transmission at any instant of time. It can be argued that the availability of significant white space spectrum might pose a challenge in its implementation with 5G, however, recent surveys in technologically mature locations have yielded excessive amount of white spaces in the TV band. During a study in Japan, more than 100 MHz of TVWS has been observed in about 84.3% of the country's area [24]. In USA, about 79% of the spectrum is unused whereas this figure is 56% for the European region [25]. For rural areas, the numbers for unused spectrum are likely to boost further. Therefore, it would be unnecessary to apply the cognitive mobile virtual network operator (C-MVNO) model for spectrum sensing and leasing, which only works best for non-extreme sensing available probabilities [26].

In our 5G-TVWS architecture, MBS would behave as a macrocell gateway and would render the UBS as a rural gateway. This rural gateway node would act as a lynchpin to feed the UHF network for facilitating Internet access to rural side. Figure 1 illustrates that UBS is deployed at the same cell-site as that of a MBS. Due to the absence of apriori knowledge of prevalent signal, Energy Detection method [27] appears as the appropriate technique for spectrum sensing with additional benefit of low computation complexity, considering the low user density in the rural area. Although it might not differentiate between noise and signal as a result of threshold comparison when pitted against Cyclostationary Feature Detection, it would still be the preferred option for minimizing the collision probability with primary user of the spectrum and to ensure communication with minimal interference.

A geographically central location in small villages can be earmarked as Village Connectivity Center (VCC) which would be used for connectivity with rural gateway (UBS). Since the transmission is being carried out on TV band frequencies in the UHF band, therefore, a conventional terrestrial TV antenna may be utilized as UHF receiver for connectivity with UBS. This UHF receiver would be located at the VCC point.

After the Internet has been extended to VCC points within a village, the last mile solution has to be simple and economical with reasonable range, so that a significant population may be able to benefit from the Internet. Subramanian et al. have suggested the use of WiLD (Wi-Fi over Long Distance) links as a promising solution especially for low user density regions [28]. The problem can be solved by using WiLD links, albeit it would require costly and specialized equipment towards user end. However, the complexity of link setup and issues pertaining to its stability over long periods of time render it unfit for rural use [29]. A comparative analysis of various technologies that can be considered for rural deployment is summarized in Table II.

TABLE II
COMPARATIVE ANALYSIS OF WIRELESS TECHNOLOGIES [10], [29] – [32]

| Technology | Range (KM) | Throughput | Rural Concern |
|---|---|---|---|
| Wi-Fi | 0.05-0.45 | 600 Mbps | Range |
| WiLD | 100-280 | 3-4 Mbps | Stability, Cost |
| WiMAX | 0.3-49 | 35-70 Mbps | Cost, Complexity |
| Satellite | Not related | 5-25 Mbps | Costly equipment |
| TVWS (UHF) | 10-30 | 2-45 Mbps | Spectrum sensing |

A more appropriate approach would be to design a local wireless cluster by creating a Wi-Fi Access Point (AP) in order to resolve the last mile access in sparsely populated areas. Although there can be various choices for last mile connectivity, Wi-Fi is preferred since it is cheaper than other options and does not require licensed spectrum for operation. Moreover, user equipment such as laptop/tablet can directly connect to the Wi-Fi AP without the need for any additional hardware. On a similar note, small rural settlements/hamlets can be clustered so that they may be encompassed within a single VCC Point coverage area. Therefore, TV white spaces in UHF spectrum may be utilized for backhauling data from VCC sites to UBS. The Wi-Fi Access Point would also be housed at the VCC site.

Depending upon the geographical topology, these VCC sites can be functionally categorized as (i) Exclusive Access Points and (ii) Relay Points. Exclusive Access Points are those VCC sites which comprise of UHF receiver and Wi-Fi transmitter, and are rendering services exclusively as a Wi-Fi access point. On the other hand, those VCC locations which behave as relay points can be viewed as the enhanced version of Exclusive Access Points. Relay Point would comprise a UHF transmitter in addition to a UHF receiver (for connectivity with rural gateway) and Wi-Fi transceiver. Wi-Fi transceiver would take care of the population in the close vicinity and UHF transmitter would be used to relay the data to far-off VCC points, which fall outside the coverage radius of the rural gateway UBS.

Figure 2 exemplifies how the various VCC points can be categorized as Exclusive Access Points and Relay Points within this architecture. It may be noted that every VCC is basically a single Access Point, which may be accordingly classified further. The scenario depicted comprises of six villages (A to F) which are fed through a single UHF base station collocated with MBS. The Relay Point has been formulated to provide connectivity to those areas (villages B and C) which lie outside the coverage radius of the rural gateway UBS. Villages D and E have been clustered since they both can be fed by smart deployment of a single Wi-Fi Access Point. Moreover, same VCC site is also being used to extend the Internet to far-off village cluster (B and C), so it is not functioning exclusively as an Access Point, therefore we term it as a Relay Point. In terms of functionality, it can be seen that Relay Point is a combination of Exclusive Access Point and UBS.

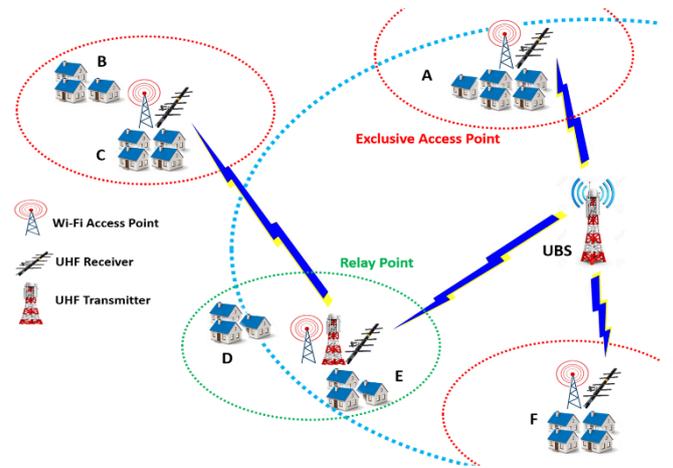

Fig. 2: Depiction of Exclusive Access Points and Relay Points for TVWS Connectivity to Rural Areas

V. COST CONSIDERATIONS FOR TVWS

The cost considerations in terms of CAPEX and OPEX have been deliberated separately and are elaborated in ensuing paragraphs.

*A. CAPEX Analysis*

CAPEX encapsulates the cost of radio BSs required for coverage of a certain area along with the construction costs of the cell site. The parameter considered for CAPEX analysis is the cost of infrastructure per user. It is worth mentioning that the platform/construction costs of the UBS cell site have not been considered since this amount is already catered in the MBS CAPEX for urban connectivity in the original 5G rollout plan.

The analysis assumes a certain rural territory of effective area $A$ having $N$ number of users. The effective area $A$ is in fact service area of the network which is being covered through numerous access points, therefore it may also extend outside the coverage radius of the main UBS. Other parameters used for CAPEX analysis are listed in Table III.

TABLE III
CAPEX PARAMETERS

| Parameter | Description |
|---|---|
| $c_U$ | Cost of a UBS |
| $c_A$ | Cost of an exclusive access point |
| $c_R$ | Cost of a relay point |
| $n_A$ | No of exclusive access points |
| $n_R$ | No of relay points |
| $\lambda$ | User density per unit area |
| $R$ | Coverage radius of access point |

It may be noted that relay point functions both as a UHF base station (UBS) as well as a Wi-Fi access point of that particular location where it is deployed. Since both functionalities require different hardware altogether, therefore it is safe to conclude that its cost is sum of the two. i.e.

$$c_R \approx c_A + c_U$$

Now, total number of VCCs in service area $A$ of a Gateway BS is given by

$$n_A + n_R = \left\lceil \frac{A}{\pi R^2} \right\rceil$$

Above equation can be written as

$$n_A + n_R = \left\lceil \frac{N}{\lambda \pi R^2} \right\rceil$$

Total incurred infrastructure cost for a single gateway BS can be calculated as

$$C_{infra} = c_U + c_A n_A + c_R n_R$$

$$C_{infra} = c_U + c_A n_A + c_A n_R + c_U n_R$$

$$C_{infra} = c_A(n_A + n_R) + c_U(n_R + 1)$$

$$C_{infra} = \frac{c_A N}{\lambda \pi R^2} + c_U(n_R + 1)$$

Cost per user is then given by

$$C_{infra/user} = \frac{c_A}{\lambda \pi R^2} + \frac{c_U(n_R + 1)}{N}$$

$$C_{infra/user} = \frac{1}{N}\left(\frac{c_A A}{\pi R^2} + c_U(n_R + 1)\right)$$

From above CAPEX representation, following inferences can be deduced.

- The infrastructure cost per user is most sensitive to the coverage radius of the Wi-Fi Access Point. If this cost is to be reduced further, it would require an increase in coverage radius. On the other hand, increase in coverage area is only possible at the prospect of an expensive Wi-Fi Access Point. So lower cost per user would necessitate a tradeoff between these two parameters.
- The use of relays would increase the infrastructure cost per user as illustrated in Figure 3. The MBS sites located at the suburbs of a city would be of maximum utility in our proposal. In case if the location of a certain village is such that it can be covered both by a relay as well as by installation of a new UBS at closest MBS site, operators would prefer UBS installation owing to similar cost effects to that of a relay. This would also ensure that in case of point of failure at any node/access point, minimum number of users are affected. Moreover, it would take care of network capacity as well due to the available bandwidth constraints. On the other hand, having no relays in the network may also lead to coverage holes, therefore it would be a cost-based tradeoff between number of relay points and UBS sites for thorough coverage since we are preferring availability over performance.

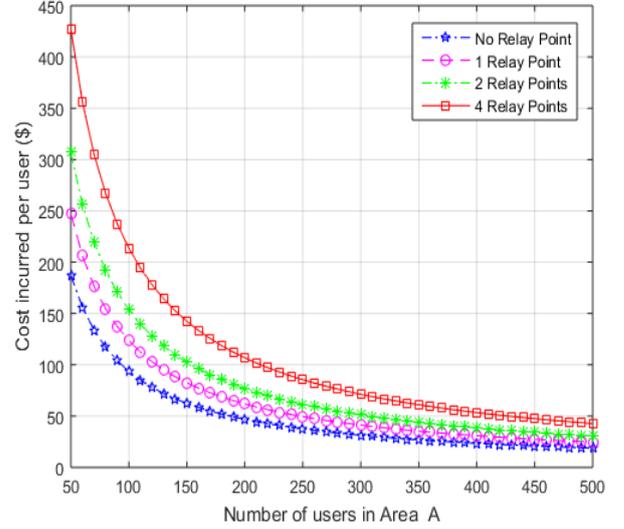

Fig. 3: Effect of Relay Points on Cost per User for 5G-TVWS Case

In light of above, this situation can be formulated into a non-linear optimization (Mixed Integer) problem for minimizing the infrastructure cost per user in the following manner:

Minimize

$$\frac{1}{N}\left(\frac{c_A A}{\pi R^2} + c_U(n_R + 1)\right)$$

subject to

$$n_A + n_R - \frac{A}{\pi R^2} \geq 0$$

$$n_A - n_R \geq 0$$

$$c_A + c_U - c_R = 0$$

$$n_A, n_R, R, c_A, c_R \geq 0$$

Cost of relay point would rise with increase in the cost of an exclusive access point, so the minimization problem comprises of five variable entities. Depending upon the geographical territory and area span of rural structure, the optimum number of relays and exclusive access points can be calculated by this approach. It is worth mentioning that this methodology is only applicable to those cases in which 5G is being used as backhaul instead of conventional options, where operators utilize optical fiber or other wireless alternatives for backhauling of network traffic. The use of fiber optics as backhaul requires the additional costs of excavation, cable laying, cable costs and site rentals as well in addition to specialized end equipment for inter-conversion between two different media; fiber and UHF.

TABLE IV
ESTIMATED COST OF INFRASTRUCTURE ITEMS

| Item | ~Estimated Cost ($) |
|---|---|
| UHF transmitter | 2000-3000 |
| Platform construction/Mast/Tower | 800 |
| Spectrum Database Manager | 1000 |
| Wi-Fi transceiver | 1000 |
| UHF receiver | 100 |
| TV UHF Band Device | 650 |
| Optical Fiber (per KM) | 15000 |

The generic costs associated with key infrastructure entities are listed in Table IV. It is worth mentioning that the costs highlighted are assumptions estimated from various references [18], [33], [34] and might vary to a limited extent in terms of market value. Figure 4 shows how rural Internet connectivity can be made practicable with 5G-TVWS backhaul in comparison with conventional approaches, where optical fiber is used. The cost effects for using TVWS with 1 KM fiber and 3 KM fiber are almost 2.5 times and 5 times respectively to that of using 5G as backhaul. In some cases, the practical lengths for far off villages might require up to 50 KM optical fiber in conventional solutions, which will shoot the infrastructure cost exponentially. In this way, these results demonstrate why the concept of TVWS slotted with conventional backhaul approaches could not be materialized on a large scale yet and how it may become the desirable rural connectivity solution when dovetailed with 5G network.

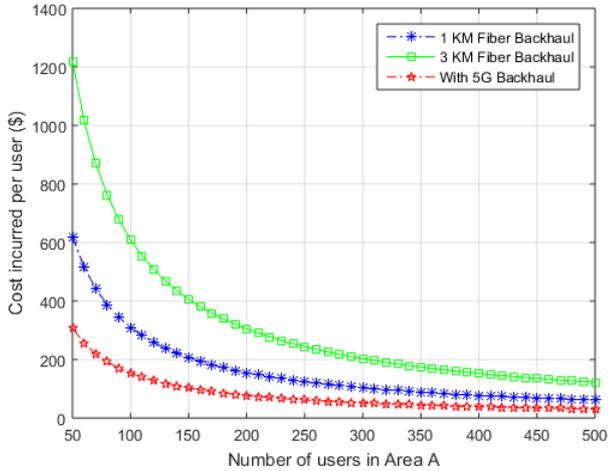

Fig. 4: TVWS Deployment Costs against Various Backhaul Approaches

### B. OPEX Analysis

Although CAPEX costs may be managed through optimization, similar strategy may be adopted to cut down the recurring costs for availability of power. OPEX costs in this case can be taken care of in two major ways:

*1) Power Saving Schedule.* Since most of the rural population depends upon agriculture for a living, therefore the switching ON/OFF mechanism of the exclusive access points can be formulated on the basis of comparative study of the traffic profile, when there is no data traffic (e.g. later half of the night etc).

*2) Use of Solar Panels.* In general, the grid/power supply in rural areas is not as reliable as in urban areas, therefore use of renewable energy sources can be considered. Since most of the rural areas are located closer to equator as compared to north/south poles, therefore the use of solar panels can be considered a viable option for these places. The size of solar panels would be based on the reliability extent of available power and average data traffic over time. A comprehensive analysis on operating costs with the help of energy scheduling algorithms has been carried out in [35] and may be considered for this case. However, energy scheduling algorithms may increase the system complexity as we are targeting a rural population with low user density, therefore, we dwell on the cost minimization scenario comprising of solar panel and national grid. The availability of battery as backup would also be needed in case of power failure after daytime. The defined parameters are listed in Table V and the quantity in parentheses represents time instant.

TABLE V
POWER PROVISIONING PARAMETERS

| Parameter | Description |
|---|---|
| $E_g/E_b/E_p$ | Energy stored in Grid/Battery/Solar Panel |
| $E_x$ | Energy stored in X (X is symbolic) |
| $E_{x-y}$ | Energy transferred from X to Y (X,Y are symbolic) |
| $L$ | Energy consumed by Load |
| $\rho$ | Battery Charging Efficiency |
| $\varphi$ | Battery Discharging Efficiency |
| $c_g$ | Cost of Grid power per consumption unit |
| $c_p$ | Cost of Solar Panel per unit area |
| $A_p$ | Area of Solar Panel |
| $\gamma$ | Solar Panel Efficiency per unit area |
| $\sigma$ | Input Solar Energy |

We consider a scenario where electric power is available at a VCC site from national grid. Since the electricity available in rural areas might have stability concerns, so we also consider the case of using a battery which may be charged for subject use. The battery would have some charging and discharging efficiency, which is essential for realistic analysis of cost concerns. To cater for power failures, the solar panel would also be needed as a backup source. The energy can be stored in battery via national grid as well as solar panel. Therefore, the energy expressions for grid and solar panel can be stated as

$$E_g(t) = E_{g-l}(t) + E_{g-b}(t)$$
$$E_p(t) = E_{p-l}(t) + E_{p-b}(t)$$

Minimum energy consumed by the load of VCC is

$$L(t) = E_{g-l}(t) + E_{p-l}(t) + \varphi E_{b-l}(t)$$

In order to minimize the operating costs, we can formulate it into a linear optimization problem. i.e.

Minimize

$$\sum_{t \in T} c_g(t) E_g(t) + c_p A_p + c_b E_{b(\max)}$$

subject to
$$E_g(t) \geq E_{g-l}(t) + E_{g-b}(t)$$
$$E_p(t) \geq E_{p-l}(t) + E_{p-b}(t)$$
$$E_b(t) \leq \rho E_{g-b}(t) + \rho E_{p-b}(t) - E_{b-l}(t) + E_b(t-1)$$
$$L(t) \leq E_{g-l}(t) + E_{p-l}(t) + \varphi E_{b-l}(t)$$
$$E_p(t) = \gamma A_p \sigma(t)$$

where $E_{b(\max)}$ is the maximum capacity of the battery. Depending on the varying prices of solar panel/grid/battery, the solution of the optimization problem provides minimum cost required for the operation. Moreover, the affordability index of solar panel vs. grid/battery from vendor's viewpoint would vary for different countries worldwide.

## VI. CONCLUSION

In this paper, we have weighed various options to explore the feasibility of providing Internet access to rural population using 5G network. Due to low user density in sparsely populated areas, rural population is not dependent on Internet as compared to urban inhabitants, so they would not be encouraged to reap its benefits at high prices. Therefore, a cost effective solution has been presented in order to minimize the digital divide between rural and urban areas, which would pave the way for an advanced communication structure in villages. We have also analyzed that various existing solutions proposed by the researchers cannot be made practicable due to cost constraints, since this dual-phase problem is least likely to be solved by a one-stage solution. From an implementation viewpoint, main challenge of this approach would be to sustain good network performance in those areas which are although fed through a single gateway base station installed at the suburbs of a city, but also comprise of successive relay points within the network path. Further efforts in this direction may focus on how cellular access along with the provisioning of Internet to rural community may be made possible in conjunction with 5G in a cost-effective manner with adequate performance.